\documentclass[useAMS,usenatbib,referee]{mn2e}
\usepackage[cp866]{inputenc}
\usepackage[english]{babel}
\usepackage{graphicx}
\usepackage{amsmath}
\usepackage{epsf}

\def\la{\; \raise0.3ex\hbox{$<$\kern-0.75em\raise-1.1ex\hbox{$\sim$}}\;}
\def\ga{\;  \raise0.3ex\hbox{$>$\kern-0.75em\raise-1.1ex\hbox{$\sim$}}\;}

\def\pFn{p_{\raise-0.3ex\hbox{{\scriptsize F$\!$\raise-0.03ex\hbox{\rm n}}}}
}  % p_Fn
\def\pFp{p_{\raise-0.3ex\hbox{{\scriptsize F$\!$\raise-0.03ex\hbox{\rm p}}}}
}  % p_Fp
\def\pFe{p_{\raise-0.3ex\hbox{{\scriptsize F$\!$\raise-0.03ex\hbox{\rm e}}}}
}  % p_Fe
\def\pFmu{p_{\raise-0.3ex\hbox{{\scriptsize F$\!$\raise-0.03ex\hbox{\rm
$\mu$}}}} }  % p_Fe
\def\m@th{\mathsurround=0pt }
\def\eqalign#1{\null\,\vcenter{\openup1\jot \m@th
   \ialign{\strut$\displaystyle{##}$&$\displaystyle{{}##}$\hfil
   \crcr#1\crcr}}\,}
\newcommand{\Del}{\mbox{$\Delta$}}             % \Delta
\title[Minimal models of cooling neutron stars with accreted envelopes]
{Minimal models of cooling neutron stars with accreted envelopes}
\author[A.~D.~Kaminker, M.~E.~Gusakov, D.~G.~Yakovlev, O.~Y.~Gnedin]
{ A.~D.~Kaminker$^{1}$\thanks{E-mail:
kam@astro.ioffe.ru (ADK), gusakov@astro.ioffe.ru (MEG),
yak@astro.ioffe.ru (DGY), ognedin@astronomy.ohio-state.edu (OYG)}, 
          M.~E.~Gusakov$^{1}$,     
          D.~G.~Yakovlev$^{1}$,          
          and
          O.~Y.~Gnedin$^{2}$\\
$^{1}$Ioffe Physical Technical Institute,
Politekhnicheskaya 26, 194021 Saint-Petersburg, Russia\\
$^{2}$Ohio State University,
Department of Astronomy, 140 W 18th Avenue,
Columbus, OH 43210, USA
}
\begin{document}
\date{Accepted 2005 xxxx. Received 2005 xxxx; 
in original form 2005 xxxx}

\pagerange{\pageref{firstpage}--\pageref{lastpage}} \pubyear{2005}

\maketitle
\label{firstpage}

\begin{abstract}
We study the ``minimal'' cooling scenario of superfluid neutron stars 
with nucleon cores,  where the direct Urca process is forbidden 
and the enhanced cooling is produced by the neutrino 
emission due to Cooper pairing of neutrons. 
Extending our previous consideration 
(Gusakov et al.\ 2004a), we 
include
the effects 
of accreted envelopes of light elements.
We employ phenomenological
density-dependent critical temperatures
$T_{\rm cp}(\rho)$ and $T_{\rm cnt}(\rho)$ of singlet-state proton
and triplet-state neutron pairing in a stellar core, as well as
the critical temperature $T_{\rm cns}(\rho)$  
of singlet-state neutron pairing in a stellar crust. 
We show that
the presence of 
accreted envelopes
simplifies the interpretation
of observations of thermal radiation from isolated neutron stars
in the scenario of Gusakov et al.\ (2004a)
and widens the class of models for nucleon superfluidity
in neutron star interiors consistent with the
observations.
\end{abstract}


%%%%%%%%%%%%%%%%%%%%%%%%%%%%%%%%%%%%%%%%%%%%%%%%%%%%%%%%%%%%%%%%%%%%%%%%%%%%%
%**************** Section 1 ******************************
\section{Introduction}
\label{introduction}
%%%%%%%%%%%%%%%%%%%%%%%%%%%%%%%%%%%%%%%%%%%%%%%%%%%%%%%%%%%%%

New observations of thermal radiation from 
isolated middle-aged neutron stars  
(e.g., Pavlov, Zavlin \& Sanwal 2002, 
Pavlov \& Zavlin\ 2003)
initiated further development of the cooling
theory of these objects.
Its main aim is to interpret the data and
constrain still poorly known
properties of dense matter in neutron star cores,
such as the composition, the equation of state and
nucleon superfluidity (e.g., 
Yakovlev \& Pethick 2004, Page et al.\ 2004
and references therein).    

It is well-known (e.g., Yakovlev \& Pethick 2004) 
that theoretical models 
of non-superfluid neutron stars which
possess nucleon cores and cool via
the modified Urca 
process of neutrino emission
cannot explain the observations. Some neutron
stars (e.g., RX J0822--4300 and PSR B1055--52) 
are much warmer than predicted by these theories, 
while others (e.g., the Vela pulsar 
or the compact source in CTA 1) 
are much colder. 
Warmest objects can be treated 
as relatively low-mass neutron stars 
with strong proton 
(e.g., Kaminker, Haensel \& Yakovlev 2001)
or neutron (e.g., Gusakov et al.\ 2004b)      
pairing in their cores. 
Strong pairing  
suppresses the modified Urca process and 
makes the stars warmer.
Coldest stars should have higher neutrino emission than
the emission provided by the modified Urca process.
They are usually treated as massive neutron stars 
which cool either
via the powerful direct Urca process in nucleon (or nucleon/hyperon)
matter or via similar processes in kaon-condensed,
pion-condensed, or quark matter in their inner cores.

Recently Page et al.\ (2004) and Gusakov et al.\ (2004a)
proposed new scenarios of neutron star cooling
which involve only standard physics
of neutron star interiors. The neutron star cores
are assumed to contain nucleons (no exotic forms of matter)
with the forbidden
direct Urca process. Some enhancement of the
cooling can be provided by neutrino emission due
to Cooper pairing of nucleons. Page et al.\ (2004)
called their cooling scenario the ``Minimal Cooling Model''
(for its simplicity). We will also use this very properly
chosen name for the scenario of Gusakov et al.\ (2004a)
that is based on the same assumptions (but 
differs in their realization; see below).
    
According to our previous paper
(Gusakov et al.\ 2004a) 
the enhanced cooling 
is produced
by the neutrino emission due to Cooper pairing of neutrons
in the cores of massive neutron stars, 
while warmest
objects are thought to be
low-mass stars 
with strong proton pairing in 
%the
their 
cores. 
We assumed a phenomenological model of strong density-dependent 
singlet-state proton pairing 
%,
with the critical
temperature 
$T_{\rm cp} (\rho)$
that has the maximum value 
$T_{\rm cp}^{\rm max} \ga 5.0 \times 10^9$~K. We also
assumed a phenomenological
model of moderate triplet-state neutron pairing 
%($T_{\rm cnt} (\rho)$) 
$T_{\rm cnt} (\rho)$ 
with the maximum critical temperature 
$T_{\rm cnt}^{\rm max}  \sim  6.0 \times 10^8$~K 
shifted to higher
$\rho$, where proton pairing 
dies out. We were able to 
interpret
all the data but under
stringent constraints on the density
dependence of $T_{\rm cnt} (\rho)$.   

The present paper extends our previous analysis.
We use the same equation of state of 
matter in neutron star interiors
(Douchin \& Haensel\ 2001)
and the same model of triplet-state neutron pairing.
However, in addition,
we take into account the effects of surface 
layers of light (accreted) elements (H and/or He), 
as well as singlet-state neutron pairing 
$T_{\rm cns}(\rho)$ in the stellar crust. 
The effects  
of accreted envelopes 
allow us to lower 
proton pairing
($T_{\rm cp}^{\rm max} \ga 10^9$~K)
required to explain the data.
This weaker proton pairing is 
consistent with recent   
microscopic calculations    
of proton critical temperatures  
by Zuo et al.\ (2004) and 
Takatsuka \& Tamagaki\  (2004) 
(although some other calculations predict much
stronger proton pairing; e.g., 
Lombardo \& Schulze 2001; 
also see references in Yakovlev, Levenfish \& Shibanov 1999,
and a recent paper by Tanigawa, Matsuzaki \& Chiba 2004).  

Let us emphasize the difference of cooling scenarios of
Page et al.\ (2004) and Gusakov et al.\ (2004a).
In particular, Page et al.\ (2004) 
used several selected models of triplet-state
neutron pairing provided by microscopic theories. 
Corresponding
cooling curves do not depend sensitively on neutron star
mass and do not allow the authors to explain all
the data in the frame of one physical model of neutron
star interiors. 
In contrast, Gusakov et al.\ (2004a) used phenomenological
models of triplet-state pairing and succeeded to explain
all the data (although under stringent constraints
on these models; see their paper for details). 
Note that Page el al.\ (2004) analyzed
the effect of accreted envelopes
on their minimal cooling models but our models are
different and require separate analysis. Our main aim
is to interpret all the data assuming the same
physics (equation of state and superfluid properties)
in the interiors of all neutron stars.

%%%%%%%%%%%%%%%%%%%%%%%%%%%%%%%%%%%%%%%%%%%%%%%%%%%%%%%%%%%%%%%%%%%%%%%%%%%%
\section{Observations}
\label{observations}
%%%%%%%%%%%%%%%%%%%%%%%%%%%%%%%%%%%%%%%%%%%%%%%%%%%%%%%%%%%%%%%%%%%%%%%%%%%%

Table 1 summarizes observations of isolated
neutron stars,
whose thermal surface radiation has been detected or constrained.
We present the estimated stellar age $t$,
the effective surface temperature $T_{\rm s}^\infty$
and the surface thermal luminosities
$L_{\rm s}^\infty$ (as detected by a distant observer).
%%%% They would move the figures here.
%Observational limits of $T_{\rm s}^\infty$ and
%$L_{\rm s}^\infty$ are represented in Figs.~1--4
%and Fig.~3, respectively.
The data on $t$ and $T_{\rm s}^\infty$ 
are described by Gusakov et al.\ (2004a) 
in more detail, with two exceptions.
First, following Slane et al.\ (2004a), 
we slightly lower the upper limit
on the surface temperature $T_{\rm s}^\infty$
of PSR~J0205+6449 in the supernova remnant 3C~58
($T_{\rm s}^\infty< 1.02$~MK  instead of $1.1$~MK). 
Second,  we include into consideration the 
central X-ray source RX J0007.0+7303 in the supernova remnant CTA~1. 

For PSR~J0205+6449 we adopt the age of the historical
supernova SN~1181 ($t\approx$820 yr).
However notice, that recently Chevalier (2004, 2005) 
presented arguments in favor for a larger age of the pulsar wind
nebula in 3C~58 ($t=$2400$\pm 500$~yr).
Were this the actual age of the neutron star, its 
interpretation would be easier.

For the source RX J0007.0+7303 we adopt the age of its host 
supernova remnant CTA~1 (G119.5+10.2).    
According to Slane et al.\ (2004b), 
the age is $t = 13$~kyr.  
Following Halpern et al.\ (2004) 
we assume the neutron star age limits $10$~kyr $\la t \la 30$~kyr. 
As for RX~J0205+6449,  the
Crab pulsar and RX J0007.0+7303,
no thermal radiation component has
been detected from these objects, and  
only the upper limits on 
$T_{\rm s}^\infty$ have been set (Slane et al.\ 2004a,
Weisskopf et al.\ 2004, Slane et al.\ 2004b,
Halpern et al.\ 2004).

%%%%%%%%%%%%%%%%%%%%%%%%%%%%%%%%%%%%%%%%%%%%%%%%%%%%%%%%%%%%%%%%%%%%%%
\renewcommand{\arraystretch}{1.2}
\begin{table*}%[!t]   % "*" ignores the twocolumn-format if adopted
\caption[]{Observational limits on surface temperatures and thermal
luminosities of isolated
neutron stars}
\label{tab:observ}
\begin{center}
\begin{tabular}{ l  c  c  c  l c }
\hline
\hline
Source & $t$ [kyr] & $T_{\rm s}^\infty$ [MK] &  Confid.\  & 
Refs.$^{c)}$ & $\lg L_{\rm s}^\infty$ [erg/s]  \\
\hline
\hline
PSR J0205+6449 (in 3C 58)    & 0.82    & $<$1.02$^{~b)}$  & 99.8\%     & 
S04a   & $<33.29$  \\
PSR B0531+21 (Crab)        &    1    & $<$2.0$^{~b)}$   &  99.8\%    & 
W04 & $<34.45$  \\
RX J0822--4300     & 2--5    & 1.6--1.9$^{~a)}$ & 90\% & 
ZTP99 & 33.9--34.2  \\
%                                              Zavlin, Tr\"umper, Pavlov 1999
1E 1207.4--5209        & 3--20 & 1.4--1.9$^{~a)}$ & 90\% & 
ZPS04 & 33.67--34.20  \\
%                                              Zavlin, Pavlov, Tr\"umper 1998
RX J0007.0+7303  (in CTA 1)  & 10--30  &     %%$\sim$13  
$<$ 0.66$^{~b)}$ & --   &
H04 & $<32.54$  \\ 
PSR B0833--45 (Vela)    & 11--25  & 0.65--0.71$^{~a)}$ & 68\% & 
P01 & 32.19--32.67  \\
%                                Pavlov, Zavlin,  Sanwal 2003 Bad Honnef
PSR B1706--44       & $\sim$17 & 0.82$^{+0.01}_{-0.34}$$^{~a)}$ & 68\% &
M04 & 31.66--32.94  \\
%                                Pavlov, Zavlin,  Sanwal 2003 Bad Honnef
PSR J0538+2817       & $30 \pm 4$ & $\sim 0.87$$^{~a)}$ & -- &
ZP04 & 32.32--33.33  \\
%PSR B0656+14       &  $\sim$110 & 0.53$^{+0.04}_{-0.03}$$^{~a)}$  & --  &
%Anderson et al.\ (1993) \\
PSR B0633+1748 (Geminga)    & $\sim$340 & $\sim 0.5$$^{~b)}$ & -- &
%Zavlin \& Pavlov (2004) \\
K05 & 31.34--32.37  \\
RX~J1856.4--3754     & $\sim$500 & $<$0.65  & -- & 
G04a & $<32.5$  \\
PSR~B1055--52       & $\sim$540 & $\sim$0.75$^{~b)}$ & -- &
PZ03 & 32.05--33.08  \\
RX J0720.4--3125   & $\sim 1300$ & $\sim 0.51$ %$^{~a)}$  
& -- &
MZH03 & 31.37--32.40  \\
\hline
\end{tabular}
\end{center}
\small{
\begin{flushleft}
$^{a)}$ Inferred using a hydrogen atmosphere model\\
%\end{flushleft}
%\begin{flushleft}
$^{b)}$ Inferred using the black-body spectrum\\
%\end{flushleft}
%\begin{flushleft}
$^{c)}$ 
S04a -- Slane et al.\ (2004a);
W04 -- Weisskopf et al.\ (2004);
ZTP99 -- Zavlin, Tr\"umper \& Pavlov (1999); 
ZPS04 -- Zavlin, Pavlov \& Sanwal (2004);
H04 -- Halpern et al.\ (2004);
P01 -- Pavlov et al.\ (2001);
M04 -- McGowan et al.\ (2004);
ZP04 -- Zavlin \& Pavlov (2004);
K05 -- Kargaltsev et al. (2005);
G04a -- see Gusakov et al.\ (2004a);
PZ03 -- Pavlov \& Zavlin (2003);
MZH03 -- Motch, Zavlin \& Haberl (2003)
\end{flushleft}
}
\end{table*}
%%%%%%%%%%%%%%%%%%%%%%%%%%%%%%%%%%%%%%%%%%%%%%%%%%%%%%%%%%%%%%%%

The surface temperatures of some %youngest
sources from Table~1 (labeled by $^{a)}$)
have been obtained by fitting their thermal radiation
spectra with hydrogen atmosphere models. Such models are
more consistent with other information on these sources
(e.g., Pavlov \& Zavlin  2003) 
than the blackbody model. For other sources 
(e.g., for the Geminga pulsar and PSR B1055--52,
labeled by $^{b)}$) we present the values
of $T_{\rm s}^\infty$ inferred using the blackbody spectrum
because this spectrum is more consistent for these 
sources. The surface temperature of RX J1856.4--3754 
is still uncertain. Following Gusakov et al.\
(2004a) we adopt the upper
limit $T_{\rm s}^\infty < 0.65$~MK. Finally, $T_{\rm s}^\infty$
for RX J0720.4--3125 is taken from Motch et al.\  
(2003), who interpreted the observed spectrum 
with a model of a hydrogen atmosphere of finite depth.             
Note also the new results by Kargaltsev et al.\ (2005) for Geminga
presented in Table~1. 
These authors confirm the observational value of
$T_{\rm s}^\infty$ reported by 
Zavlin \& Pavlov (2004). Taking into account
systematic uncertainties of $T_{\rm s}^\infty$  discussed
by Kargaltsev et al.\ (2005)  we retain 
%(in Figs.~1--4) 
$20\%$ errorbars   
adopted by Gusakov et al.\ (2004a)
and erroneously referred 
to $90\%$ confidence level in their Table~1. 
Following Gusakov et al.\ (2004a), the same $20\%$ errorbars
will be adopted for PSR J0538+2817, PSR B1055--52, and RX J0720.4--3128.

As noted by several authors (e.g., Page et al.\ 2004),
it may be instructive to compare the cooling theory with
measured values of 
stellar thermal surface luminosities $L_{\rm s}^\infty$, 
rather than with $T_{\rm s}^\infty$.
The data on $L_{\rm s}^\infty$ 
%associated with the observational 
%values on $T_{\rm s}^\infty$
are also collected
in Table~1. The luminosity is related to the effective surface
temperature via
\begin{equation}
     L_{\rm s}^\infty = 4 \pi \sigma R_\infty^2 (T_{\rm s}^{\infty})^4,
\label{Luma}
\end{equation}
where $\sigma$ is the Stefan-Boltzmann constant,        
$R_\infty = R/\sqrt{1-2GM/(c^2 R)}$ is the so called
apparent radius of a neutron star (as would be detected
by a distant observer if a telescope could resolve the star),
$R$ is the circumferential radius, and $M$ the gravitational stellar
mass. Thus, the luminosity is determined by the effective
temperature and neutron star radius; an uncertainty
in $L_{\rm s}^\infty$ is produced by uncertainties in $T_{\rm s}^\infty$
and $R_\infty$. 
We have already described the values of $T_{\rm s}^\infty$.
As for the values of $R_\infty$, we 
vary them (with two exceptions indicated below)
within the reasonable theoretical interval for
neutron star radii, $R_\infty$=11--16~km;
%%%%%
%All values of $L_{\rm s}^\infty$
%are expressed in erg~s$^{-1}$; 
%%%%%
while translating $R$ into $R_\infty$ we always set
$M=1.4\,M_\odot$. 

%Let us outline the data on $L_{\rm s}^\infty$ in Table~1.
In Table~1
the upper limits on $L_{\rm s}^\infty$
%%%%%%%%
%$\lg L_{\rm s}^\infty<33.29$, 
%$\lg L_{\rm s}^\infty<34.45$,   
%$\lg L_{\rm s}^\infty<32.54$, and
%$\lg L_{\rm s}^\infty<32.5$
%%%%%%%%%
for PSR J0205+6449,
the Crab pulsar,
RX J0007.0+7303, and RX J1856.4--3754
%respectively, 
were obtained assuming
$R_\infty=16$~km. 

The luminosities 
%%%%%%%%
%$\lg L_{\rm s}^\infty=$33.9--34.2 
%and 33.67--34.20 
%%%%%%%%
of RX J0822--4300 and
1E 1207.4--5209 have been calculated from the values of $T_{\rm s}^\infty$
obtained by Zavlin et al.\ (1999) and Zavlin et al.\ (2004),
respectively. We have taken the same fixed radius $R=10$~km ($R_\infty=13$~km) 
which was used by the cited authors 
to fit the observed
spectra with the hydrogen atmosphere models.
All other values of $L_{\rm s}^\infty$ in Table~1
have been obtained 
by varying $R_\infty$ within the interval $R_\infty=$11--16~km.

The central values of $L_{\rm s}^\infty$ have been calculated 
taking into account the central values of $T_{\rm s}^\infty$
from
Table~1 and the values of $R$ (or $R_\infty$) 
obtained in cited papers from spectral fits,
except for the Vela pulsar, where we 
set $R_\infty=13$~km.  
For PSR B1706--44, PSR J0538+2817,
and RX J0720.4--3125 these values of $R$
have been taking 
12~km, 10.5~km, and 10~km, as suggested
by McGowan et al.\ (2004), Zavlin \& Pavlov\ (2004), and 
Motch et al.\  (2003), respectively.   
For the Geminga pulsar we have used the value $R=10.6$~km 
from Zavlin \& Pavlov\ (2004), and for PSR B1055--52
%the central point of $L_{\rm s}^\infty$ has been obtained
%taking 
we set $R=13$~km from Pavlov \& Zavlin\ (2003). 

In all the cases, the limits of $L_{\rm s}^\infty$ presented in Table~1
seem to be rather uncertain. Although, in principle, 
the luminosities $L_{\rm s}^\infty$
can be measured/constrained more accurately than $T_{\rm s}^\infty$
(by exact measuring the distance and the bolometric thermal flux), 
it is not so for the sources collected in Table~1 mainly
due to large uncertainties in measured distances to the sources
(see, e.g., Page et al.\ 2004).
Nevertheless, comparing observed and 
theoretical luminosities 
of cooling neutron stars
seems to be useful. 
Our limits of $L_{\rm s}^\infty$ are in reasonable
agreement with corresponding limits given by Page et al.\ (2004).
%%%%%%%%%%%%%%%%%%%%%%%%%%%
The main differences refer to 
the Geminga pulsar and  
1E 1207.4--5209. In the first case
the limits of $L_{\rm s}^\infty$ 
presented by Page et al.\ (2004)
correlate with
too low apparent radius of the star, 
$R_\infty<6$~km, for the temperature limits 
adopted in their paper.  
%%%%%%%%%%%%%%%%%%%%%%%%%%%%%%
In the second case
Page et al.\ 
%adopted  
used
$L_{\rm s}^\infty = L_{\rm s} \,(R/R_\infty)^2$  
with $L_{\rm s}=5.0^{+4.3}_{-1.8} \times 10^{33}$~erg s$^{-1}$
from Zavlin, Pavlov \& Tr\"umper\ (1998). 
The value of $L_{\rm s}$ was
possibly underestimated by Zavlin et al.\ (1998),
because their value of
$T_{\rm s}$ 
was indicated later by Zavlin et al.\ (2004)   
as $T_{\rm s}^\infty$.
     
Also, let us note that the radii of our neutron star models 
used for the cooling calculations presented below
are consistent with the radii
used for the interpretation of the data.

%%%%%%%%%%%%%%%%%%%%%%%%%%%%%%%%%%%%%%%%%%%%%%%%%%%%%%%%%%%%%%%%%%%%%%%%%%%
\section{Physics input and calculations}
\label{physics}
%%%%%%%%%%%%%%%%%%%%%%%%%%%%%%%%%%%%%%%%%%%%%%%%%%%%%%%%%%%%%%%%%%%%%%%%%%%
 
The cooling calculations have been done
using our general relativistic 
cooling code (Gnedin, Yakovlev \& Potekhin 2001). 
At the initial  
cooling stage ($t \la 100$ yr)
the main cooling mechanism is the neutrino emission but
the stellar interior stays highly non-isothermal. 
At the next  
stage ($10^2$~yr $ \la t \la 10^5$ yr) the neutrino emission
is dominant but the stellar interior is isothermal.  
Later ($t \ga 10^5$ yr)  
the star cools predominantly through the surface photon emission.

Following Gusakov el al.\ (2004a) we adopt the 
moderately stiff equation of state for the neutron star matter 
suggested by Douchin \& Haensel (2001). In this case
a neutron star core (a region of density
$\rho > 1.3 \times 10^{14}$ g~cm$^{-3}$) 
consists of neutrons with the admixture
of protons, electrons and muons. All constituents
exist everywhere in the core, except for muons
which appear at $\rho> 2.03 \times 10^{14}$~g~cm$^{-3}$. 
The most massive stable
star has the (gravitational) mass $M=M_{\rm max}=2.05\ M_\odot$,
the central density 
$\rho_{\rm c}=2.9 \times 10^{15}$~g~cm$^{-3}$, 
and the (circumferential) radius $R=9.99$~km.  
The parameters of neutron stars with some other masses are given by
Gusakov et al.\ (2004a).

\begin{figure*}%[t]
%\centering
\epsfysize=80mm
\epsffile[18 145 569 418]{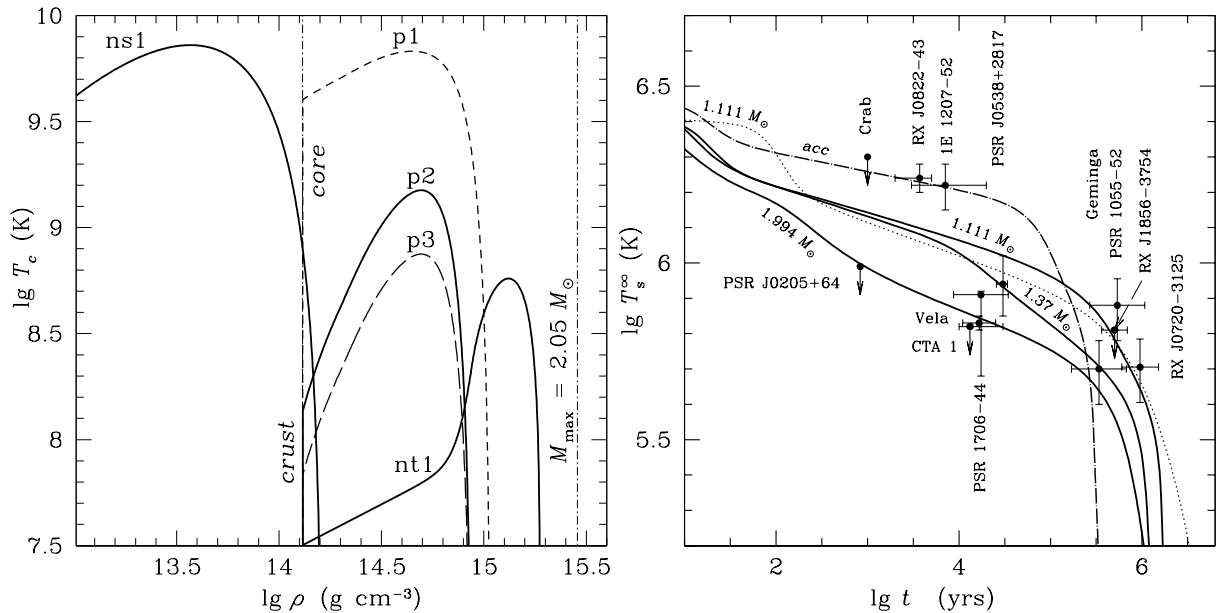}
\caption{
{\it Left:} \, 
Density dependence of critical temperatures 
for three models p1, p2, and  p3 
of singlet-state proton pairing and 
one model nt1 of triplet-state neutron 
pairing in the neutron star 
core, as well as model ns1 of singlet-state neutron pairing in
the stellar crust.
Vertical dot-and-dash lines indicate the crust-core boundary
and the central density of the maximum-mass neutron star.  
{\it Right:} \, 
Observational limits of surface temperatures 
of neutron stars (Table 1) 
as compared with theoretical cooling curves.  
Three solid curves refer
to neutron stars of different masses (indicated near the curves)
with nucleon pairing p2, nt1, 
and ns1. 
The dot-and-dashed curve refers to an $M=1.111 M_{\odot}$
neutron star with the same   
superfluidity, but with the 
accreted envelope  
of the mass $\Del M=10^{-8}\,M$. 
The dotted curve is for a
non-superfluid star of  
the same mass without any accreted envelope.       
}
\label{fig1}
\end{figure*}

The employed equation of state 
forbids the powerful direct Urca
process of neutrino emission (Lattimer et al.\ 1991)
in all stable neutron
stars ($M \leq M_{\max}$). 
Accordingly, a non-superfluid neutron
star of any mass 
in the range $M_\odot \la M \leq M_{\rm max}$
(without any accreted envelope)
will have almost the same (universal) cooling curve
$T_{\rm s}^\infty(t)$ (the dotted curve
in the right panel of Fig.\ \ref{fig1}).
At the neutrino cooling stage, this curve is determined
by the modified Urca process and is almost independent
of the equation of state in the stellar core 
(see, e.g., Yakovlev \& Pethick 2004 and references therein). 
As seen from Fig.\ \ref{fig1}, this universal cooling curve  
cannot explain the data. 
We will show that all the data can be explained
assuming nucleon superfluidity in the internal layers
of neutron stars and the  
presence of accreted envelopes (of light elements).

Following the standard procedure    
(Gudmundsson, Pethick \& Epstein  1983) 
our code calculates
heat transport in the neutron-star interior 
($\rho>\rho_{\rm b}=10^{10}$ g~cm$^{-3}$) and uses the
predetermined relation between  
the effective surface temperature
$T_{\rm s}$  and the temperature $T_{\rm b}$ at the 
bottom of the surface heat-blanketing envelope ($\rho < \rho_{\rm b}$). 
We use the relation calculated
by Potekhin, Chabrier, \& Yakovlev (1997) and
updated by Potekhin et al.\ (2003). We will employ the models 
of blanketing envelopes made of iron 
(which is the standard assumption) and envelopes
containing light elements. 

The detailed description of these models 
is given by Potekhin et al.\ (2003). 
The thermal energy in the heat-blanketing envelope 
is mainly conducted by electrons.
The thermal conductivity of electrons which scatter off
lighter ions in the accreted envelope
is higher than the conductivity in the iron envelope. 
This means that the accreted
envelope is more heat transparent than the iron one, resulting in  
higher  $T_{\rm s}$ for the same $T_{\rm b}$.
This rise of the surface temperature
depends on $T_{\rm b}$ and $\Delta M$,
the mass of light elements (hydrogen and/or helium,
with a possible carbon/oxygen layer at the bottom of the
accreted envelope as a result of nuclear burning of lighter elements).
Potekhin et al.\ (1997, 2003) varied the boundaries of layers
containing different elements within physically reasonable limits
and found that the resulting relation between $T_{\rm s}$ and $T_{\rm b}$
is remarkably insensitive to these variations and depends mainly
on $\Delta M$. However, $\Delta M$ cannot exceed $\sim 10^{-7}\,M$,
because at higher $\Delta M$ the bottom 
density of the accreted envelope would
exceed $10^{10}$ g~cm$^{-3}$. At such high densities, light elements 
(including carbon/oxygen) would
rapidly transform into heavier ones. 

At the neutrino cooling stage $T_{\rm b}$ is governed
by the neutrino emission from the stellar interior and
is almost independent of conductive properties in
the heat-blanketing envelope. In contrast,
at the photon cooling stage the star with the
accreted envelope has lower 
$T_{\rm b}$ and, consequently, lower 
$T_{\rm s}$ 
due to higher heat transparency
of the surface layers. This leads to faster photon cooling
through the surface 
(for not too cold stars; see, 
e.g., Potekhin et al.\ 1997).   

The cooling of a neutron star
is sensitive to superfluidity of nucleons in the stellar core
and to superfluidity of free neutrons in the inner stellar 
crust. Any superfluidity
is characterized by its own density-dependent critical 
temperature $T_{\rm c}(\rho)$.
Microscopic theories predict mainly 
(i) singlet-state ($^1$S$_0$) pairing of 
neutrons ($T_{\rm c}=T_{\rm cns}$) in the inner crust 
and the outermost core;
(ii) $^1$S$_0$ proton pairing in the core ($T_{\rm c}=T_{\rm cp}$); 
and (iii) triplet-state ($^3$P$_2$) neutron pairing in the core 
($T_{\rm c}=T_{\rm cnt}$).   
These theories give a large scatter of critical temperatures,
from $\sim 10^{10}$~K to $\sim 10^8$~K and lower, depending on
a nucleon-nucleon interaction model and a many-body theory
employed (e.g., Lombardo \& Schulze\ 2001,
Yakovlev et al.\ 1999;
also see recent papers by
Schwenk \& Friman\ 2004, 
Takatsuka \& Tamagaki\ 2004, 
Zuo et al.\ 2004, Tanigawa et al.\ 2004).
Because of these huge theoretical uncertainties,
we will not rely on any specific microscopic results  
but will treat $T_{\rm cp}(\rho)$ and $T_{\rm cn}(\rho)$
as phenomenological functions of $\rho$ 
(which can be varied in physically reasonable limits).
Our aim will be to constrain these functions
by comparing theoretical cooling
curves with the observations.

Superfluidity of nucleons 
affects the heat capacity  
and suppresses neutrino processes
such as 
Urca and nucleon-nucleon bremsstrahlung processes 
(as reviewed, e.g., by Yakovlev et al.\ 1999). 
It also introduces 
an additional neutrino emission
mechanism associated with Cooper pairing of 
nucleons (Flowers, Ruderman \& Sutherland 1976). 
All these effects
of superfluidity are incorporated into our cooling code.
While calculating the neutrino emission due to
Cooper pairing of protons we use 
phenomenological values of weak-interaction parameters 
renormalized by many-body effects (the same as
in Gusakov et al.\ 2004b). 

In the left panel of Fig.\ \ref{fig1}
we plot models for nucleon pairing adopted
in our calculations:  
one model ns1 of strong singlet-state
pairing of neutrons (with the peak
of  $T_{\rm cns}(\rho)$ approximately equal to
$T_{\rm cns}^{\rm max} \approx  7 \times 10^9$~K); 
three models of proton pairing --  
strong p1, moderately strong p2, and 
moderate p3  
($T_{\rm cp}^{\rm max} \simeq 6.8 \times 10^9$~K,
$1.5 \times 10^9$~K, and $7.5 \times 10^8$~K, respectively);
and one model nt1 of moderate
triplet-state neutron pairing   
($T_{\rm cn}^{\rm max} \sim 6 \times 10^8$~K).
Models p1 and nt1   
are the same as in Gusakov et al.\ (2004a). 
Now we add
models of weaker proton pairing 
(particularly, p2). Strong proton pairing has been
predicted in a number of publications (e.g., Tanigawa et al.\ 2004)
while other publications predict much weaker proton pairing
(e.g., Zuo et al.\ 2004, Takatsuka \& Tamagaki 2004).

%%%%%%%%%%%%%%%%%%%%%%%%%%%%%%%%%%%%%%%%%%%%%%%%%%%%%%%%%%%%%%%%%%%%%
%%%%%%%%%%%%%%%%%%%%%%%%%%%%%%%%%%%%%%%%%%%%%%%%%%%%%%%%%%%%%%%%%%%%%%%%%

As seen from the right panel
of Fig.\ \ref{fig1},  
proton pairing p2 
combined with strong crustal 
superfluidity of neutrons ns1
%is insufficient 
results in too cold low-mass neutron stars.
The neutrino emission 
due to Cooper pairing of protons 
in the core and of neutrons in the inner crust  
(see Section \ref{low-mass}) 
accelerates the cooling and does not allow us to 
explain the observations of the young and hot neutron stars, 
RX J0822--4300 and 1E 1207.4--5209.
However, this cooling scenario 
is consistent with 
the observations of the old 
and warm neutron stars, 
PSR 1055--52 and RX J0720.4--3125.
Accreted envelopes
can rise the surface temperatures  
of middle-aged neutron stars and 
explain the observations 
of RX J0822--4300 and 1E 1207.4--5209.
This is demonstrated by
the dot-and-dashed  cooling curve for the low-mass star
with the accreted envelope of the mass $\Delta M = 10^{-8} M$. 

%%%%%%%%%%%%%%%%%%%%%%%%%%%%%%%%%%%%%%%%%%%%%%%%%%
Our interpretation of the neutron stars coldest
for their age (PSR J0205+6449 in 3C 58, 
RX J0007.0+7303 in CTA~1, the Vela 
and Geminga pulsars) remains the same as in Gusakov et al.\ (2004a). 
These objects can be treated 
as massive neutron stars ($M \ga 1.9 \,M_\odot$)
with moderate triplet-state neutron
pairing nt1 in their inner cores      
where proton pairing p1 (as well as p2 and p3)  
dies out (the left panel of 
Fig.\ \ref{fig1}). Our phenomenological pairing model nt1
seems specific (shifted to too high densities $\rho$).
However similar models have been obtained from
microscopic theories (e.g., see the curve $m^*=0.73$
in Fig.\ 1 of Takatsuka \& Tamagaki 1997).

In this way we come to the same three distinct 
classes of cooling neutron stars as in  
Gusakov et al.\ (2004a) 
(and generally as in
Kaminker et al.\ 2002).  
The first class contains low-mass stars 
whose surface layers are composed either of iron 
or of light elements
(solid or dot-and-dashed cooling 
curves, respectively, for the $M=1.111\ M_\odot$ star 
in Fig.\ \ref{fig1}). 
Another class contains high-mass stars which show
{\it enhanced}
cooling (the solid curve for the $M=1.994\ M_\odot$ star)
produced by the
neutrino emission due to Cooper pairing of neutrons.
Finally, there is a class of medium-mass neutron stars 
(the solid curve for the $M=1.37\  M_\odot$ star) which show
intermediate cooling.  Their cooling curves
fill in the space between the upper curve for low-mass stars  
and the lower curve for high-mass stars. These curves
explain the observations of PSR B1706--44, PSR J0538+2817,
and RX J1856.4--3754.  

%%%%%%%%%%%%%%%%%%%%%%%%%%%%%%%%%%%%%%%%%%%%%%%%%%%%%%%%%%%%%%%%%%%%%%%%%%%
\section{Cooling of low-mass neutron stars}
\label{low-mass}
%%%%%%%%%%%%%%%%%%%%%%%%%%%%%%%%%%%%%%%%%%%%%%%%%%%%%%%%%%%%%%%%%%%%%%%%%%%

As has been shown in Section \ref{physics}, 
the presence of light elements
on the surfaces of the younger and hotter
neutron stars, RX~J0822-4300 and 1E~1207.4-5209,
can allow us to explain their observations 
if we assume moderately strong proton pairing p2 in their interiors.
This pairing is also consistent with the observations of 
the old and warmest sources, 
PSR B1055--52 and RX J0720.4--3125. We interpret
all these sources as low-mass neutron stars.
Let us analyze
the main cooling regulators of such stars.

%%%%%%%%%%%%%%%%%%%%%%%%%%%%%%%%%%%%%%%%%%%%%%%%%%%%%%%%%%%%%%
\begin{figure*}
\centering
\epsfysize=80mm
\epsffile[18 145 569 418]{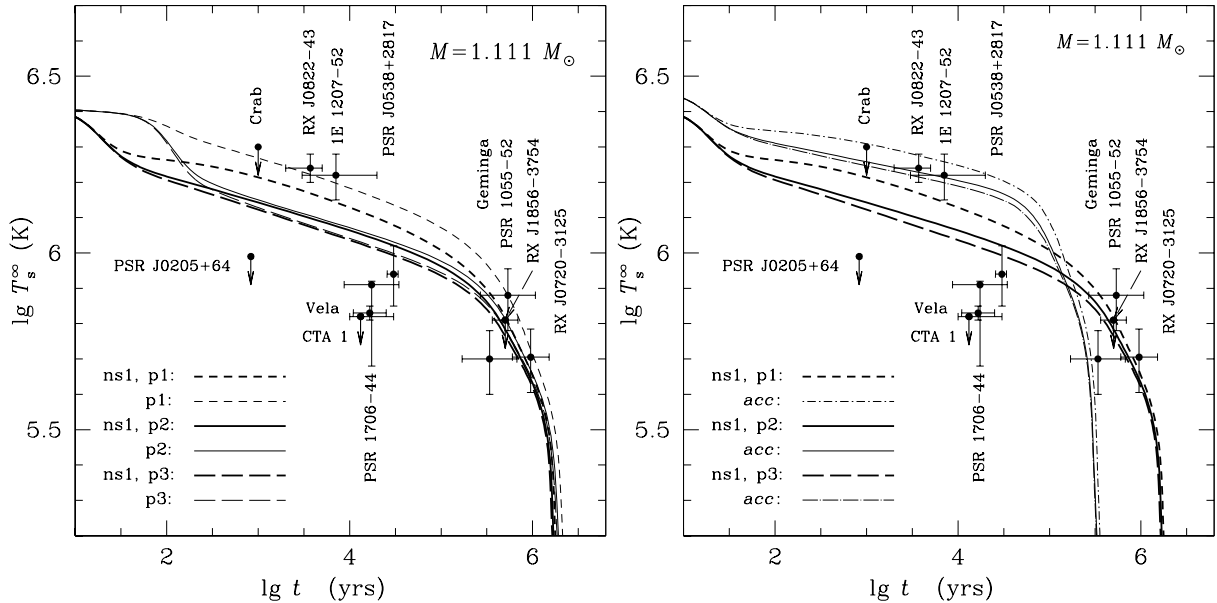}
\caption{ 
 Cooling of low-mass (1.111 $M_{\odot}$) neutron stars
 with proton pairing (models p1, p2 or p3)
 and neutron pairing nt1
 (weak at low densities, Fig.\ \protect{\ref{fig1}})
 in the stellar core
 (thin short-dashed, solid, and long-dashed curves, respectively),
 with the aim to interpret the observations of neutron stars
 hottest for their age (RX J0822--4300, 
 1E 1207.4--5209, PSR B1055--52, and RX J0720.4--3125).    
 Thick curves include, in addition, 
 the effects of crustal neutron superfluidity ns1.
 {\it Left:} \, No accreted envelopes.
 {\it Right:}  \, 
 Thick curves are the same as in the {\it left} panel. 
 Thin dot-and-dashed and solid 
 curves include, in addition, the effects of accreted envelope 
 of the mass $\Del M= 10^{-8}\,M$.  
 }
\label{fig2}
\end{figure*}
%%%%%%%%%%%%%%%%%%%%%%%%%%%%%%%%%%%%%%%%%%%%%%%%%%%%%%%%%%%%%%%
 
In our case, triplet-state neutron pairing
in low-mass stars is weak.
For the adopted equation of state
of Douchin \& Haensel (2001), this implies
$T_{\rm cn}(\rho) \la 2 \times 10^8$ K at $\rho \la 8 \times 10^{14}$
g~cm$^{-3}$. Under this condition, neutron pairing does
not affect the cooling of low-mass
stars ($M \la 1.1\,M_\odot$) at least at the neutrino cooling stage.  
The thin short-dash line in the left panel of Fig.\ \ref{fig2} shows that 
(in the absence of crustal pairing)
strong proton pairing p1 is needed to explain 
the data on all neutron stars hottest
for their age (Gusakov et al.\ 2004a). 
In contrast, cooling curves for  
moderately strong proton pairing p2 (the thin solid line) 
and moderate pairing p3 (the thin long-dashed line)  
go essentially lower than the curve for
pairing p1, being inconsistent 
with the observations of  
RX J0822--4300 and 1E 1207.4--5209. 
More rapid cooling
for these two models of proton superfluidity
is provided by
the neutrino emission due to Cooper pairing of protons
which occurs at $t \sim 50$--$100$~yr. 

Thick lines in the left panel of Fig.\ \ref{fig2}
demonstrate the additional effect of
neutron pairing ns1 in the crust.
Comparing three thick lines,
one can see that crustal neutron pairing  
noticeably accelerates only very slow cooling 
of low-mass neutron stars with   
strong proton pairing p1 in their cores
(Yakovlev et al.\ 2001, 2002).   
In that case the neutrino luminosity 
due to Cooper pairing of neutrons
in the stellar crust 
at $t \la 3 \times 10^5$~yr may dominate  
the total neutrino luminosity of the stellar core.
Moreover, at $t \ga 3 \times 10^5$~yr 
crustal neutron pairing
reduces the heat capacity of the crust. 
Both effects 
accelerate the cooling and decrease $T_{\rm s}^\infty$,
violating the interpretation of the two hottest sources,
RX J0822--4300 and 1E 1207.4--5209.  
Any model of weaker crustal superfluidity 
will only
bring cooling curves closer to thin ones
and simplify the interpretation of the 
observations.

On the other hand,
for moderately strong (p2) or moderate (p3)  
proton pairing in the core, 
the effects of strong 
crustal neutron pairing  
on the cooling of middle-aged neutron stars 
($10^3$~yr $\la t \la 10^5$~yr)
are almost negligible.  
The neutrino emission
due to crustal Cooper pairing of neutrons   
can noticeably accelerate the cooling 
and decrease $T_{\rm s}^\infty$ only during the  
internal thermal relaxation stage ($t \la 100$~yr). 
  
The right panel of Fig.\ \ref{fig2} 
demonstrates that the observations of  
RX J0822--4300 and 1E 1207.4--5209 
can be explained by adopting any 
model of proton pairing (p1, p2 or p3),
model ns1 of crustal superfluidity, and
the presence of an accreted  envelope of the 
mass $\Delta M = 10^{-8}\,M$
(thin lines). 

Let us note, that 
the upper dot-and-short-dashed cooling curve  
goes higher than is needed to interpret
the observations 
of the young and hottest source
RX J0822--4300.  Accordingly,  
following Yakovlev et al.\ (2002)
(also see Potekhin et al.\ 2003), 
we may assume the presence of a thinner accreted envelope 
(e.g., $\Delta M \sim 10^{-11} M_\odot$) 
to interpret the observations of RX J0822--4300 and 1E 1207.4--5209
(for the combination of p1 and ns1 pairing).
The stronger proton core superfluidity,
the less massive accreted envelope
is needed for the interpretation of the data for these two stars. 

In order to explain the old and warmest sources, 
PSR B1055--52 and RX J0720.4--3125, 
we will treat them
as low-mass stars with the iron surface 
and proton pairing p2 in the core 
(or similar 
model of 
pairing 
with the peak of critical temperature 
$T_{\rm cp}^{\rm max} \ga 10^9$~K). 
Moreover, the presence 
of any crustal neutron pairing
(for example, ns1;
thick solid lines in Fig.\ \ref{fig2}),  
does not violate the interpretation of these sources. 
Note that proton pairing p3 
(thick long-dashed lines) 
is less appropriate for the interpretation of these  
sources than pairing p2. 
Therefore, we adopt proton pairing p2 
as the basic model for a 
new cooling scenario. Obviously, 
any model of stronger proton pairing 
(with higher $T_{\rm cp}(\rho)$)
is better consistent with the observations.
 
%%%%%%%%%%%%%%%%%%%%%%%%%%%%%%%%%%%%%%%%%%%%%%%%%%%%%%%%%%%%%%%%%%%%%%%%%%%%%%%%
\section{Accreted envelopes and cooling of neutron stars}
\label{accr}
%%%%%%%%%%%%%%%%%%%%%%%%%%%%%%%%%%%%%%%%%%%%%%%%%%%%%%%%%%%%%%%%%%%%%%%%%%%%%%%%
 
\begin{figure*}%[t]
\centering
\epsfxsize=\hsize
\epsffile[25 100 555 360]{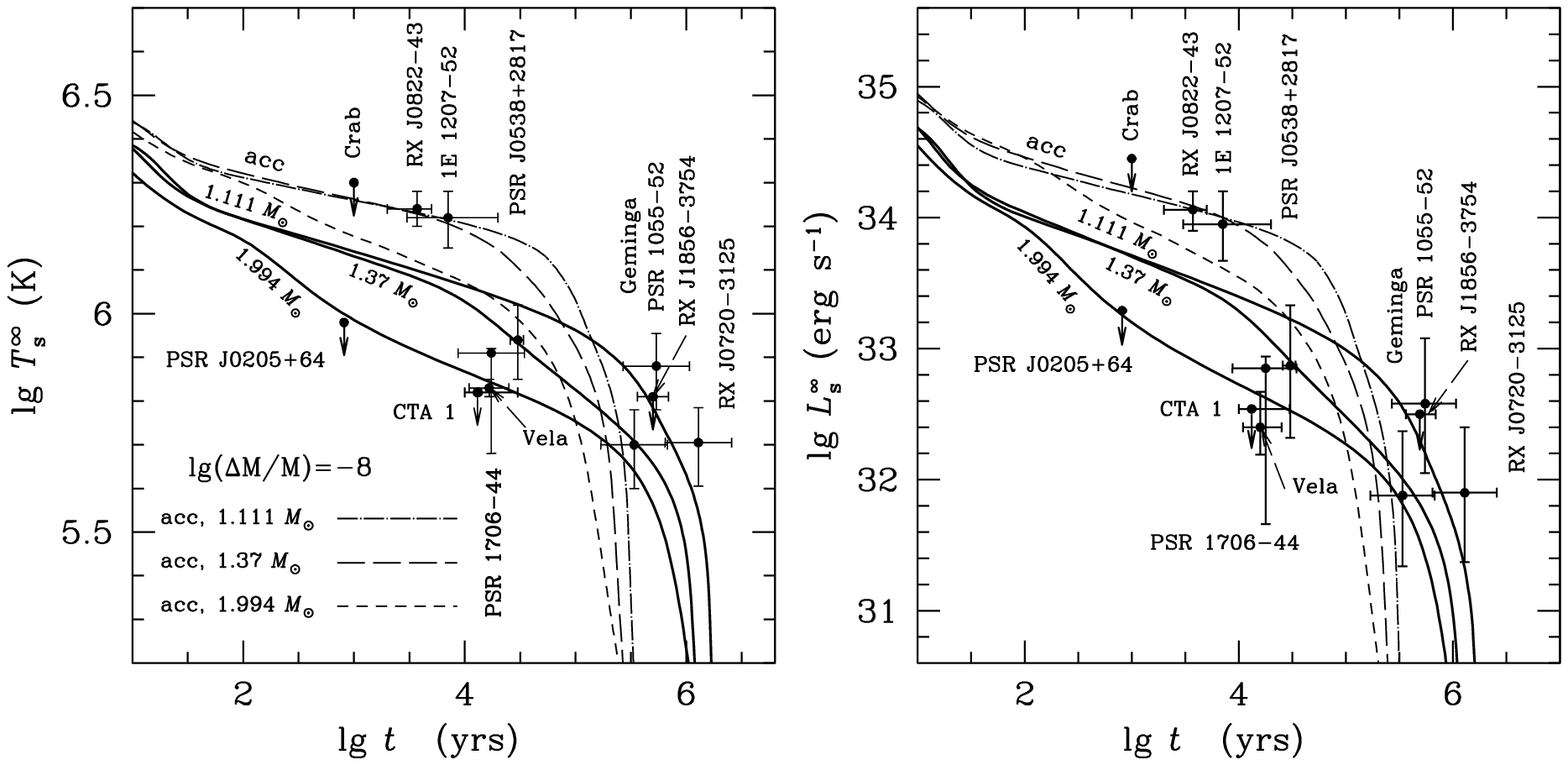}
\caption{
Cooling curves 
of neutron stars with different masses and 
nucleon pairing p2, nt1, and ns1 
(Fig.\ \protect{\ref{fig1}}) versus 
observations. Thin dot-and-dashed ($M=1.111\ M_{\odot}$), 
long dashed ($M=1.37\ M_{\odot}$), and short dashed 
($M=1.994\ M_{\odot}$) curves are calculated including 
the effects of accreted envelopes
($\Del M = 10^{-8}\,M$). 
Three thick solid curves are the 
same as the solid curves in the right panel of 
Fig.\ \protect{\ref{fig1}}. The left and right panels show
the same cooling curves but as functions $T_{\rm s}^\infty(t)$ and 
$L_{\rm s}^\infty(t)$, respectively.  
}
\label{fig3}
\end{figure*}

Figure \ref{fig3} illustrates the  
effects of the accreted envelopes of 
the mass
$\Delta M = 10^{-8}\,M$
on the cooling of neutron stars with different masses
and the same nucleon pairing 
(models p2, nt1, and ns1).
For comparison, we present also the cooling curves  
for stars with iron surface 
(thick solid lines)
and the same nucleon superfluidity 
(also see the right panel of Fig.\ \ref{fig1}). 
Note that the effect of crustal superfluidity
on the cooling of such stars is unimportant. 

%%%%%%%%%%%%%%%%%%%%%%%%%%%%%%%%%%%%%%%%%%%%%%%%%%%%%
%%%%%%%%%%%%%%%%%%%%%%%%%%%%%%%%%%%%%%%%%%%%%%%%%%%%%%%%

In the left panel of Fig.\ \ref{fig3}
we present our traditional cooling curves $T_{\rm s}^\infty(t)$
and compare them with the data on the surface temperatures.
On the right panel we show the temporal evolution
of the surface thermal luminosity $L_{\rm s}^\infty(t)$
and compare it with the data (Table~1). Both representations
of the same cooling processes are seen to be in a
reasonably good agreement
although the data on $L_{\rm s}^\infty$ are generally less certain
and seem to be currently less conclusive (because, as a rule, the luminosity
of the selected sources is determined less accurately than their
surface temperature as discussed in Section \ref{observations}). 

Figure \ref{fig3} shows a  
strong rise of cooling curves 
for neutron stars with accreted envelopes at the 
neutrino cooling stage ($t \la 3 \times 10^4$~yr) and 
their steep decrease at the photon cooling stage.
Their photon stage 
starts earlier than for stars with the iron surface.  
Assuming the presence of 
accreted envelopes, we
can explain the observations of the
young and hottest neutron stars,
RX J0822--4300 and 1E 1207.4--5209, treating them
either as low-mass or as medium-mass stars.
In contrast, the observations of the old and warmest objects,  
PSR B1055--52 and RX J0720.4--3125, 
can be explained only by treating them as low-mass stars    
with the iron surfaces  
and with moderately strong (or strong) 
proton pairing inside. 

It was shown by Chang \& Bildsten\ (2003, 2004)
that the mass of light elements may decrease with time, particularly   
due to diffusive nuclear burning. The characteristic
burning time $\tau$ can be considered as an additional
cooling regulator.  
Following  Chang \& Bildsten\ (2003, 2004)
and Page et al.\ (2004)
we assume that the mass of light elements 
decreases with time as
$\Delta M (t) = \Delta M_0 \,\exp(- t/\tau) $, where
$\Delta M_0$ is the initial mass. 
 
%%%%%%%%%%%%%%%%%%%%%%%%%%%%%%%%%%%%%%%%%%%%%%%%%%%%%
\begin{figure}%[t]
\centering
\epsfxsize=\hsize
\epsffile[18  145  570  691]{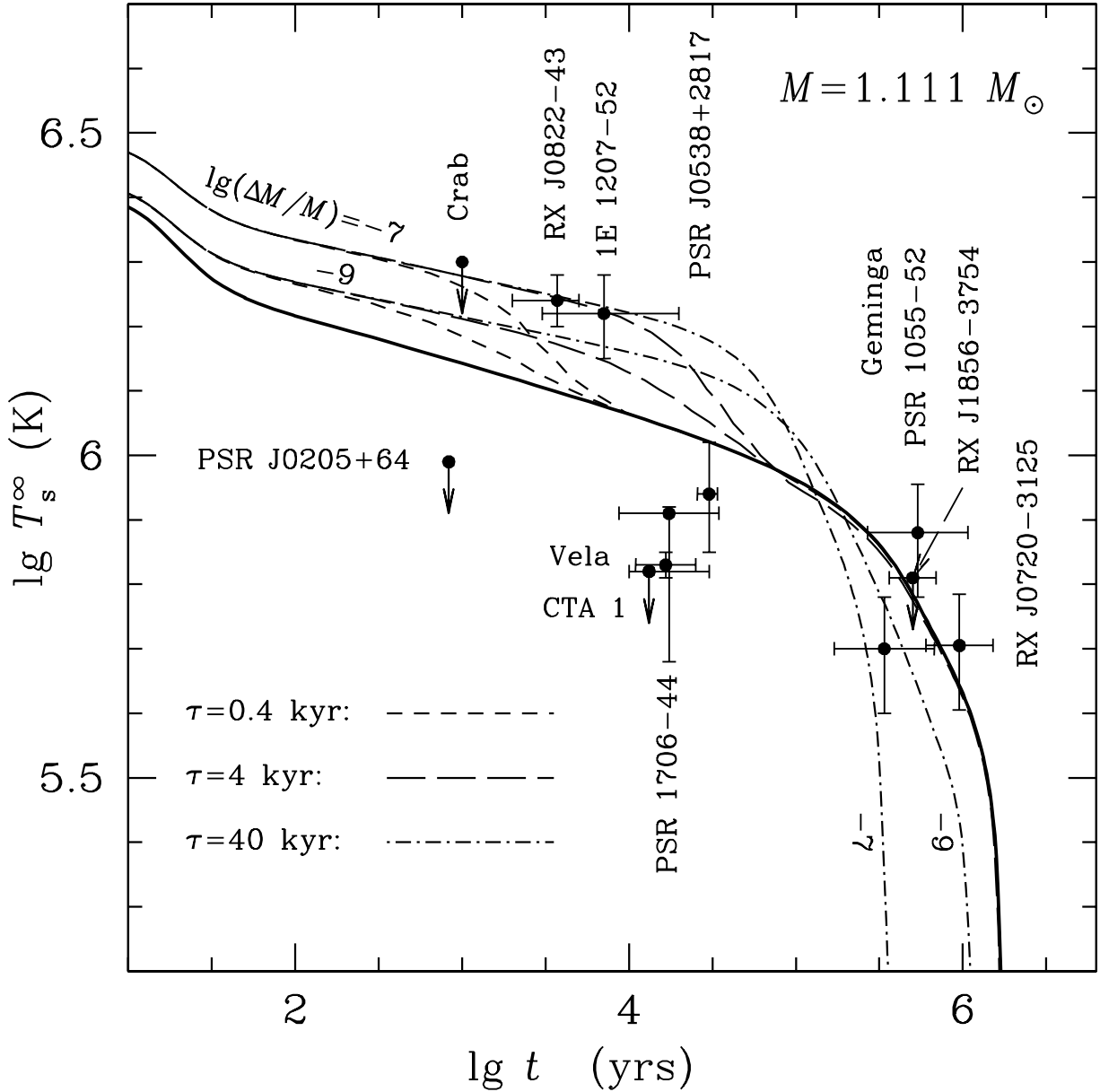}
\caption{
Cooling curves of 1.111 $M_{\odot}$ neutron stars
with nucleon 
pairing p2, nt1, and ns1
versus observations. The thick solid curve is the same as  
in Fig.~\protect{\ref{fig3}}; 
other curves are calculated assuming an exponential  
decay of accreted envelopes with characteristic
times $\tau =$ 0.4~kyr (short-dashed lines), 
4~kyr (long-dashed lines), 
and 40~kyr (dot-and-dashed lines) for two values
of the initial accreted mass,  $\Del M_0/M = 10^{-9}$ and 
$10^{-7}$.   
}
\label{fig4}
\end{figure}
%%%%%%%%%%%%%%%%%%%%%%%%%%%%%%

Figure \ref{fig4} illustrates the effect of variable
mass of the accretion envelope on the cooling of 
the $M=1.111 M_\odot$ neutron star.
All cooling curves are calculated assuming 
nucleon pairing p2, nt1, and ns1.
The thick solid line is our typical cooling 
curve for a low-mass superfluid
star without any accreted envelope. 
We use two values of the initial mass of light elements  
$\Delta M_0/ M = 10^{-9}$ and $10^{-7}$ and present thus      
three pairs of cooling curves for 
three characteristic times $\tau$.
When $\tau$ is lower   
than the time of the transition 
from the neutrino cooling 
stage to the photon stage
($\tau < 3 \times 10^4$~yr) 
we obtain a smooth transition of the cooling track
from the regime of highest temperatures in young stars
to the regime of lower temperatures in old stars
(cf.\ the curves for $M=1.111\ M_\odot$ in 
Figs.\ \ref{fig3} and  \ref{fig4}).  
This effect has been pointed out 
by Page et al.\ (2004).
At $\tau \ga 3 \times 10^5$~yr 
cooling curves merge 
into the {\it limiting}  
curve obtained for constant  
$\Delta M = \Delta M_0 $.
In the intermediate case of   
$3 \times 10^4$~yr $ \la  \tau \la  3 \times 10^5$~yr 
the cooling curves gradually approach  
this limiting curve with the increase of $\tau$. 
As seen from Fig.\ \ref{fig4},
by assuming any $\tau$ in the range
$10^3$~yr $ \la \tau \la 10^4$~yr 
one can explain the observations of all neutron stars
hottest for their age by one cooling curve. 
Note also that the value $\Delta M_0/ M = 10^{-9}$ 
is too small to explain the observations of young 
and hottest neutron stars, especially  
RX J0822--4300,  at any  $\tau$. 

As remarked by Chang \& Bildsten (2004), an accreted envelope
of a pulsar can become thinner
owing to the excavation of ions from the stellar surface
by a pulsar wind at a rate $\dot{M}\sim 2 \Omega^2 m_i\mu /ec$,
where $\Omega$ is the pulsar spin frequency, $\mu$ is the magnetic
moment, and $m_i$ is the ion mass. For an ordinary pulsar 
with the spin period $\sim 0.1$~s,
$\mu\sim 10^{30}$~G~cm$^3$, and helium surface we would
have the surface mass loss 
$\Delta M_{\rm ex} \sim 6 \times 10^{-12}\,M_\odot$  in $t \sim 10^5$ yr,
too small to affect the cooling of a star with the
initial helium layer of $\Delta M \ga 10^{-10}\,M_\odot$.
For a pulsar with much higher magnetic field and/or faster
rotation the effect may be stronger and affects the cooling.

%%%%%%%%%%%%%%%%%%%%%%%%%%%%%%%%%%%%%%%%%%%%%%%%%%%%%%%%
\section{Conclusions}
\label{concl}
%%%%%%%%%%%%%%%%%%%%%%%%%%%%%%%%%%%%%%%%%%%%%%%%%%%%%%%%

We have extended the scenario of neutron star cooling
proposed by Gusakov et al.\ (2004a)
taking into account the effects of accreted
envelopes and 
crustal singlet-state pairing of neutrons.
As stressed in Section \ref{introduction},
this scenario is different from the minimal cooling
scenario of Page et al.\ (2004). 
 
The general idea of the minimal cooling scheme   
is that the enhanced neutrino emission,
required for the interpretation of observation of neutron stars
coldest for their age, is provided by the neutrino emission
due to Cooper pairing of neutrons. 
In this case the direct Urca process or similar 
enhanced neutrino 
processes 
in kaon-condensed,
pion-condensed, or quark matter can be forbidden
in neutron stars of all masses.
 
As in Gusakov et al.\ (2004a), 
the proposed cooling scenario
imposes stringent constraints 
on the density dependence
of the critical temperature
$T_{\rm cn}(\rho)$ for triplet-state
neutron pairing in the stellar core. They 
result from the comparison
of theoretical cooling curves with the data on the three most important
``testing sources'', PSR J0205+6449,
RX J0007.0+7303, and the Vela pulsar 
(Sect.\ \ref{physics}). 
By tuning our phenomenological model
of triplet-state neutron pairing
in the stellar core
we obtain a noticeable dependence of
the cooling on neutron star mass. 
It enables us to
explain  all the data by single combination of models 
for nucleon superfluidity. 
Assuming the presence of accreted envelopes 
we obtain two additional 
parameters to regulate the cooling, which are the initial envelope mass
$\Delta M_0$  and 
its characteristic burning time $\tau$ 
(Chang \& Bildsten 2003).
  
Our interpretation implies the presence 
of moderately strong proton pairing
($T_{\rm cp}^{\rm max} \ga 10^9$~K) 
and moderate triplet-state neutron pairing
(with $T_{\rm cnt}^{\rm max} \sim 6 \times 10^8$~K)
in neutron star cores. 
Also, we have taken into account the effect of 
strong singlet-state neutron pairing
($T_{\rm cns}^{\rm max} \sim 7 \times 10^9$~K) 
in the stellar crust.
However, as shown in
Sects.\ \ref{low-mass} and \ref{accr},
the effect of crustal superfluidity
is unimportant for
cooling middle-aged neutron stars 
with moderately strong  proton  
pairing in their cores.
  
We need proton superfluidity to explain 
the observations
of the neutron stars hottest for their age. 
However, in contrast to the cooling scenario
of Gusakov et al.\ (2004a), our new
cooling scenario does not require too strong proton pairing. 
In fact, we can explain the observations of 
the old and warmest stars,  
PSR B1055--52 and RX J0720.4--3125, 
by treating them as low-mass neutron stars
(without accreted envelopes) with
moderately strong  
proton pairing in their cores.
Such phenomenological models for proton 
pairing are consistent
with recent   
microscopic calculations    
of proton critical temperatures  
by Zuo et al.\  (2004) and 
Takatsuka \& Tamagaki\  (2004). 

The young and hottest neutron stars,
RX J0822--4300 and 1E 1207.4--5209,
can also be treated  
as low-mass stars
with the same moderate proton superfluidity
in their cores but 
assuming  
the presence of accreted envelopes.   
The smaller the mass of 
the envelope,  
required for the interpretation 
of these sources,   
the stronger proton pairing 
should be assumed. 

As discussed above,
we need neutron pairing
nt1 (or similar)
to explain the observations of the stars 
coldest for their age.
However, as has been demonstrated by
Gusakov et al.\ (2004b), cooling curves 
are not too sensitive to 
exchanging neutron and proton superfluidities
($T_{\rm cp}(\rho)  \rightleftharpoons T_{\rm cn}(\rho)$)
in neutron star cores. 
Therefore, we would also be able to explain the data 
in the scenario with moderately strong neutron and 
moderate proton pairing in the stellar cores.  

Neutron star cooling can also be affected by
surface magnetic fields and by some reheating mechanisms
in neutron star interiors. We have not discussed the
effects of magnetic fields (although
they are incorporated in our cooling code).
The main reason is that these effects are weaker than
the effects discussed above (for ordinary cooling isolated
neutron stars of non-magnetar type; see, e.g.,
Yakovlev et al.\ 2002 for a detailed discussion of this point). 
Internal reheating mechanisms
(see, e.g., Page 1998a,b, and references therein),
for instance, the reheating due to the viscous dissipation
of differential rotation, are 
relatively weak and model dependent; they become
important at the photon cooling stage. No reheating
is required to explain the data in our cooling scenario.  
What is more important, that most elaborated model
equations of state of dense matter
%(Akmal, Pandharipande\ 1998)
(Akmal, Pandharipande \& Ravenhall\ 1998)
predict the operation of the direct Urca process in
most massive stable neutron stars. This should lead to
the existence of new classes of cooling neutron stars.
The scenario with the open direct Urca process
(which can be called the {\it extended minimal cooling scenario})
has been studied by Gusakov et al.\ (2005). 

It is 
important that the same physics of neutron star
interiors, which is tested by observations of isolated
(cooling) neutron stars, can also be tested   
by observations of accreting neutron stars in soft X-ray transients
(e.g., Yakovlev, Levenfish \& Haensel 2003) basing on the
hypothesis of deep crustal heating 
of such stars 
(Brown, Bildsten \& Rutledge 1998) by
pycnonuclear reactions in accreted matter (Haensel \& Zdunik 1990). 
The observations of soft X-ray
transients in quiescent states indicate 
(Yakovlev, Levenfish \& Gnedin 2005) the existence of 
rather cold neutron stars (first of all, SAX J1808.4--3658)
inconsistent with the model of neutron star structure 
proposed in the present paper.  
However, these observational indications are currently 
inconclusive (e.g., Yakovlev et al.\ 2005).
If confirmed in future observations, they could give
stronger evidence against the proposed scenario
than new observations of cooling neutron stars.
In this case the extended minimal cooling scenario
may appear to be more perspective. 
%%%%%%%%%%%%%%%%%%%%%%%%%%%%%%%%%%%%%%%%%%%%%%%%%%%%%%%%

\section*{Acknowledgements}
We are grateful to Yurii Shibanov 
for very fruitful discussions and 
critical remarks,
and to Patrick Slane for
useful discussion of the observational data.
This work was supported partly by
the RFBR (grants 03-07-90200, 05-02-22003, 05-02-16245),
the Russian Leading Science Schools (grant 1115.2003.2),
the Russian Science Support Foundation,
and by the INTAS (grant YSF 03-55-2397).

%%%%%%%%%%%%%%%%%%%%%%%%%%%%%%%%%%%%%%%%%%%%%%%%%%%%%%%%%%%

\end{document}